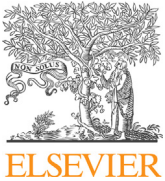
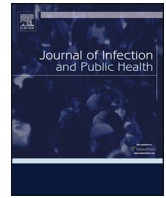

# Public sentiment analysis and topic modeling regarding COVID-19 vaccines on the Reddit social media platform: A call to action for strengthening vaccine confidence

Chad A. Melton [a], Olufunto A. Olusanya [b], Nariman Ammar [b], Arash Shaban-Nejad [a,b,*]

[a] University of Tennessee, Bredesen Center for Interdisciplinary Research and Graduate Education, Knoxville, TN, USA
[b] Center for Biomedical Informatics, Department of Pediatrics, College of Medicine, University of Tennessee Health Science Center, Memphis, TN, USA

**A B S T R A C T**

*Background:* The COVID-19 pandemic fueled one of the most rapid vaccine developments in history. However, misinformation spread through online social media often leads to negative vaccine sentiment and hesitancy.
*Methods:* To investigate COVID-19 vaccine-related discussion in social media, we conducted a sentiment analysis and Latent Dirichlet Allocation topic modeling on textual data collected from 13 Reddit communities focusing on the COVID-19 vaccine from Dec 1, 2020, to May 15, 2021. Data were aggregated and analyzed by month to detect changes in any sentiment and latent topics.
*Results:* Polarity analysis suggested these communities expressed more positive sentiment than negative regarding the vaccine-related discussions and has remained static over time. Topic modeling revealed community members mainly focused on side effects rather than outlandish conspiracy theories.
*Conclusion:* Covid-19 vaccine-related content from 13 subreddits show that the sentiments expressed in these communities are overall more positive than negative and have not meaningfully changed since December 2020. Keywords indicating vaccine hesitancy were detected throughout the LDA topic modeling. Public sentiment and topic modeling analysis regarding vaccines could facilitate the implementation of appropriate messaging, digital interventions, and new policies to promote vaccine confidence.

© 2021 The Author(s). Published by Elsevier Ltd on behalf of King Saud Bin Abdulaziz University for Health Sciences. This is an open access article under the CC BY license (http://creativecommons.org/licenses/by/4.0/).

## Introduction

In late December of 2019, the highly transmittable coronavirus disease 2019 (COVID-19) acquired through the Severe Acute Respiratory Syndrome Coronavirus 2 (SARS-COV-2), began its rampage impacting every aspect of life throughout all societies of the world. COVID-19 was declared a pandemic by the World Health Organization (WHO) in March 2020, and nearly a year later, approximately 150 million individuals have been infected (confirmed) and 2.8 million have died [1]. Vaccines are determined to be one of the most effective interventions at preventing and controlling the spread of the COVID-19 pandemic [2]. Although the accelerated development of vaccines was unprecedented, improving public sentiments for vaccine uptake and diffusing widespread skepticism towards science has been extremely challenging, particularly against the backdrop of the COVID-19 pandemic [2]. This is clearly illustrated by the unwillingness and reluctance among certain populations across the world to be vaccinated against Covid-19 [2]. Vaccine hesitancy, classified among the top ten threats to global health by the WHO, is defined as the "delay in acceptance or refusal of vaccines despite availability of vaccine services" [2]. Vaccine hesitancy/refusal/delay is perceived to originate from a diverse, multifaceted, and often concurring array of underlying factors ranging from religion, political ideology, the anti-vaccination movement, to outlandish conspiracy theories and beliefs [3]. Current drivers for COVID-19 vaccine hesitancy include disinformation, misinformation, conspiracy beliefs propagated through social media, inadequate and contradictory response from the federal government, frustrations among the general public, and fear of the unknown [4]. Apprehension regarding the vaccine's safety, side-effects, efficacy, and access also contribute to vaccine hesitancy. For instance, the recent pause in the roll-out of the Johnson & Johnson's Janssen (J&J/Janssen) COVID-

\* Corresponding author at: Center for Biomedical Informatics, Department of Pediatrics, College of Medicine, University of Tennessee Health Science Center, Memphis, TN, USA.
E-mail address: ashabann@uthsc.edu (A. Shaban-Nejad).

19 vaccine due to reported rare side effects from blood clots has reignited fears regarding vaccine uptake. Exacerbating the COVID-19 related-health messaging crises are the sensational design of vaccine-related misinformation and "fake news" which have tended to spread more rapidly than factual evidence-based information [5]. Concernedly, this spread of vaccine disinformation and misinformation ultimately leads to quantifiable negative outcomes (e.g., low vaccination rates, increasing hospitalization rates, morbidity, and mortality from vaccine-preventable diseases, etc.) [6,7]. Needless to say, the COVID-19 vaccine roll-out has been challenging due to vaccine hesitancy/delay/refusal thus the urgent need for a call to action.

As of July 15, 2021, the United States has administered about 334,000,000 doses of vaccine. Additionally, a total of 3,416,511,310 doses of vaccine have been administered worldwide (including the U.S) [1]. Despite the nebulous onslaught of disinformation and misinformation regarding the COVID-19 vaccine, results of recent surveys suggest that public opinion/confidence is improving and has increased from approximately 51–69% in the last several months [8,9]. This slight increase in positive views regarding the COVID-19 vaccine is one step in the right direction to improving vaccination acceptance. However, current vaccination rates are still significantly below the percentage threshold required for herd immunity in the U.S. (i.e., 79–90%) [10]. Therefore, it is increasingly imperative to assess public sentiment and to understand what drives vaccine hesitancy. Sentiment analysis and topic modeling are analytical tools that can be utilized to systematically identify, extract and measure the sentiment and topics from subjective textual data (e.g., obtained from social media posts, customer reviews, online survey responses, etc.) quickly, effectively, and inexpensively, as well as extrapolate common themes throughout a document.

This study seeks to examine public sentiments and opinions regarding the COVID-19 vaccine using textual data which were harvested and analyzed from Reddit (a popular social media platform). Due to recent moderately positive polling results, the hypothesis motivating our work is to investigate the sentiments reflected in discussions related to the COVID-19 vaccines throughout our data set and determine whether these sentiments reflect the public sentiment towards vaccinations Moreover, we expect to detect some evidence of vaccine hesitancy in these communities as well. During these dynamic and challenging times, we anticipate that this study will offer insight into the general public's sentiments/opinions regarding the COVID-19 vaccine using a relatively unexplored dataset. Most importantly, the results from this study will help to guide and facilitate the implementation of digital educational interventions and campaigns among vaccine-hesitant populations as well as provide additional information to public health officials to inform decision-making and policies.

**Background**

*Sentiment analysis and topic modeling of social media*

Sentiment analysis is the practice of extrapolating the sentiment of a subject, idea, event, or phenomena by computationally classifying written texts as some value of polarity (i.e., positive, negative, or neutral) [11]. Because gauging public sentiment is vastly important to determining appropriate messaging, intervention, and policies, these techniques have been used in many scientific, social, and commercial applications. Sentiment analysis of social media posts is a relatively new field. Nowadays social media data are used for a wide variety of research applications. Some early works analyzed Twitter data to detect sentiment in product reviews to inform potential consumers while others questioned and tested whether microblogs (such as Twitter) were better for sentiment analysis than longer documents [12,13]. Another study employed sentiment analysis techniques to gain insight into the 2012 U.S. Presidential election [14]. Despite many concerns related to validity, representativeness confounding and biases of social media data [15], with an estimated 3.96 billion users worldwide, social media platforms remain a valuable source of textual semantic rich data with excellent opportunities to surveil various aspects of social interaction, and especially discussions regarding public health issues.

*Natural language processing and disease surveillance*

Natural Language Processing (NLP) techniques have been used successfully in many efforts to surveil social media posts regarding vaccination and disease occurrence. Alessa and Miad (2019) monitored Twitter posts related to influenza and detected the onset of an outbreak [16]. Additionally, Raghupathi et al. (2020) showed correlations between effective public health measures and positive sentiment, as well as between increased measles infections and negative social media posts concerning the measles vaccination [17]. More recently, these techniques have been used to monitor public opinion around the world regarding mask-wearing during the pandemic [18]. Other researchers combined several machine learning classification algorithms with sentiment analysis techniques to measure public opinion around various topics related to COVID-19 [19]. Also, several sentiment analysis studies regarding the COVID-19 vaccine have been conducted as of early 2021. A study by Gbashi et al. (2021) focused on detecting the opinion of media polarity on COVID-19 vaccine in Africa with Twitter and Google News articles [20]. Furthermore, additional research investigated public sentiment in India [21,22], Indonesia [23], and China [24]. Wu et al., 2021 published a study regarding public sentiment towards the COVID-19 vaccine and topic modeling of several subreddits [25]. However, the subreddits that were chosen for analysis contained posts about the COVID-19 vaccine but were not directly related to or focused on the vaccine.

**Methodology**

*Data source*

To investigate public sentiment regarding the COVID-19 vaccine, we collected vaccine-related data from the Reddit information-sharing social media platform that is currently accessed by approximately 430 million users with approximately 50% located in the U.S. The platform is composed of user-created communities (subreddits), in which members adhere to a set of community regulations. Subreddit members have the option to post links, images, videos, and text. Community members then typically "upvote" or "downvote" a post based on their opinion of the quality of that post and/or leave comments. Depending on the distribution of votes, posts are classified as hot, new, rising, and controversial. The most popular posts within each category are then moved to the top of the community page. These comments are subjected to the same vote ranking system. The upvote/downvote system within Reddit is intended to increase the quality of the posts to minimize non-relevant material. We harvested approximately 18,000 posts from thirteen subreddits (*Vaccines, CovidVaccine, CovidVaccinated, AntiVaxxers, vaxxhappened, antivaccine, conspiracy, conspiracytheories, NoNewNormal, conspiracy_commons, COVID19, COVID,* and *coronavirus*) through the Reddit API on May 16, 2021. Because Reddit communities potentially contain some inherent bias due to strict community rules, as well as content monitoring by a moderator, these subreddits were chosen to create a non-biased dataset from a diverse selection of communities that vary widely in political views as well as position on vaccination. These subreddits were also chosen due to a large number of members (approximately five million members). Data were cleaned first by combining each subreddit into a centralized database. The data were then organized by date and then queried for terms specifically related to the COVID-19 vaccine. These terms included COVID vaccine, vaccine, vaccination, immune, immunity, COVID vaccination, corona vaccine, COVID19 vaccination, COVID-19 vaccination, coronavirus vaccination, coronavirus vaccine, COVID-19 vaccine, coronavirus vaccine, coronavirus vaccination, Moderna, Pfizer, J&J, Johnson & Johnson, COVID vax, corona vax, covid-19 vax, covid19 vax, coronavirus vax). Our finalized dataset consisted of 1401 posts and 10,240 comments (11,641 in total) written by greater than or equal to 8281 authors/users, 1048 of whom posted multiple times. In actuality, the number of authors could have been as high as 9013. These additional users are probable because Reddit removes the user ID from posts after a user deletes their account. However, the post content and upvotes remain. After data were cleaned and organized, we conducted a sentiment analysis and Latent Dirichlet Allocation (LDA) topic modeling with NLP tools in Python.

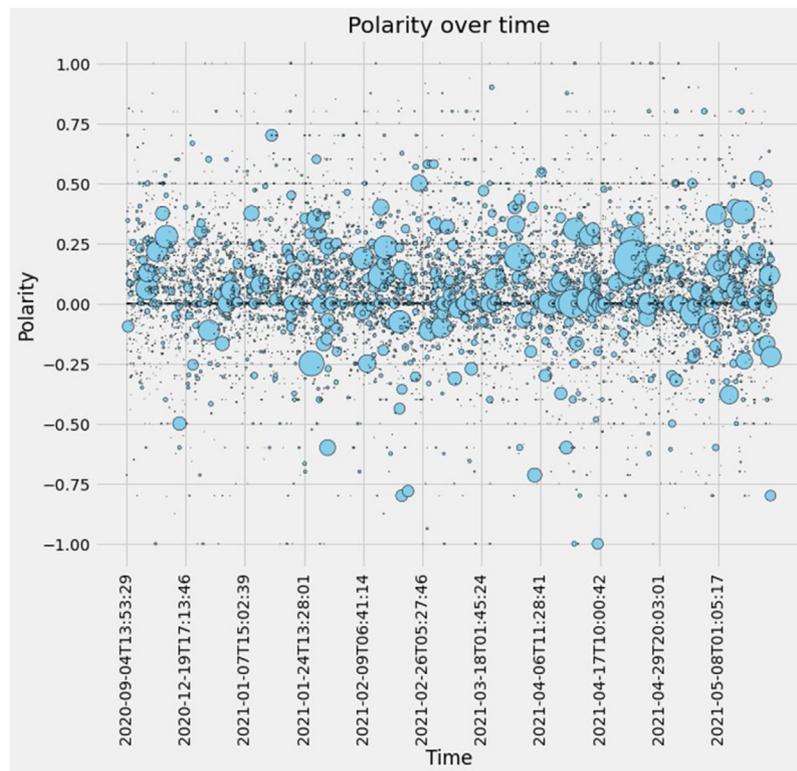

**Fig. 1.** Polarity versus time. Polarity is represented on the y-axis and time is represented on the x-axis. Data points are represented as light blue circles. Circle size indicates the number of upvotes per comment. Datapoint sizes are reflective of the vote count and are represented by larger circles and smaller quantities are represented by smaller circles. Please see https://github.com/Cheltone/NLP_Reddit for month-to-month plots.

*Analytical methods*

Our study used a lexical-based sentiment analysis. This method employs dictionaries of words with a previously assigned valence score as a reference for the text analysis. This design is somewhat similar to using labeled historical data in machine learning but computes much quicker because dictionaries have been pre-trained. In general, sentiment can be determined from several levels of complexity ranging from large volumes of text to single words or *unigram*. First, the Regex library was employed to clean and remove special characters or any remaining hyperlinks in the text of each subset. At this point, *subjectivity* and *polarity* were calculated with the TextBlob subjectivity and polarity functions. The subjectivity function returns a floating-point value between [0,1] (0 being most factual and 1 being the most opinionated). The function works by quantifying modifiers or adverbs in a sentence (e.g., *extremely lethargic*). Subjectivity values measuring between (0.4, 0.6) were classified as neutral, values greater than 0.6 were classified as "Highly Opinionated," and less than 0.4 were classified as "Least Opinionated". Polarity returns a floating-point value between [−1.0, 1.0] where −1.0 is considered to be the most negative while 1.0 is the most positive. The polarity tool works by comparing each word in a user-provided corpus to a previously defined polarity reference dictionary within the *TextBlob.sentiment.polarity* constructs [26].

The *Gensim LDAModel* algorithm was used to create LDA models for each month [27]. This technique is highly useful in detecting latent topics in large textual data. LDA assumes that documents with similar topics will use similar diction and that the topics will display a sparse Dirichlet distribution. For example, every word in the document is randomly assigned to a user-defined number of topics T. The algorithm then calculates the proportion of words in each document assigned to a topic (i.e., [p(topic T | document D)]) and then the proportion of words that were assigned to a topic over all documents (i.e., [p(word W | topic T)]). The product of these proportions is computed for each topic T and compared to every other topic T until algorithmic convergence is achieved [28]. After removing *stop words* (e.g., determiners, conjunctions, and prepositions), and lemmatizing the corpus (i.e., converting a word to its base form), coherence values were tested on 50 different LDA models to determine the most statically appropriate number of probable latent topics. Though coherence values are insightful, topics were qualitatively analyzed to double-check for content coherency rather than numerical.

Because our data set was collected from posts ranging over approximately six months, we conducted the sentiment analy-sis and LDA topic modeling using collective data ranging from December 1st, 2020 to May 15, 2021, as well as individual months. Once polarity was determined, we divided our dataset by polarity (i.e., positive, negative, neutral) and conducted further LDA based on the previously calculated polarity.

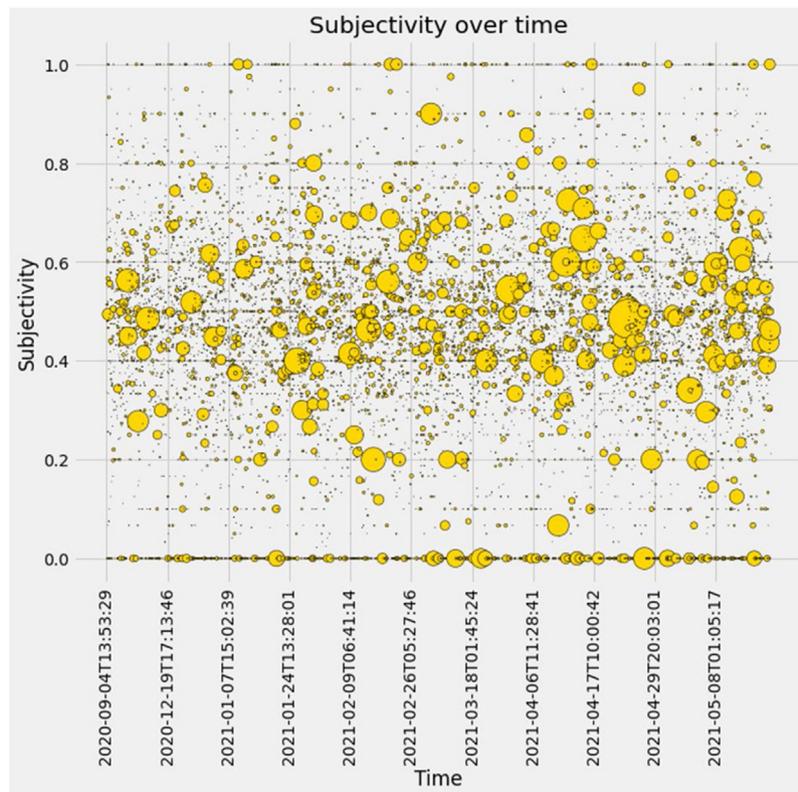

**Fig. 2.** Subjectivity versus time. Polarity is represented on the y-axis and time is represented on the x-axis. Data points are represented as blue circles. Please see https://github.com/Cheltone/NLP_Reddit for month-to-month plots.

## Results

### Combined analysis

For our combined dataset, the polarity analysis found that 56.68% of the posts measured positive, 27.69% were negative, and 15.63% neutral. The mean polarity value reported was 0.0520 and the variance was 0.0415. The subjectivity analysis reported 73.15% of the comments measured in between [0.25, 0.75] and considered neutrally subjective, 18.13% were reported to be minimally subjective (less than 0.25) while the remaining 8.72% were highly subjective (greater than 0.75). The mean subjectivity and variance were reported to be 0.4450 and 0.0560 respectively (Figs. 1 and 2). The comments from these posts received a total of 612,217 upvotes. Upvote scoring ranged from a maximum value of 11,110 upvotes, and a minimum of −135. The mean upvote count was reported at 50.04, and the mode was 7 upvotes. Comments that were classified as negative received 133,305 upvotes (23.24%). The Neutral classified comments received a total of 94,641 upvotes (16.50%). Lastly, positive classified comments received 345,607 upvotes (60.26%) (see Figs. 1 and 2).

The combined dataset LDA modeling returned an optimal number of five latent topics in our dataset. This classification was determined by computing coherence scores for 50 models. The calculation returned a high score of 0.5641 (see Figs. 3 and 4).

The LDA topic model displayed interesting results. For the collective dataset, the LDA modeling results displayed a total of five optimal latent topics. *Topics 1-4* appear to be closely related to a broader discussion of the vaccine, safety concerns, efficacy, and potential side effects. Side effects mentioned in these topics ranged from being less severe (e.g., fever, sore) to death. Keywords in these topics could suggest that these specific users consist of people that have not taken the vaccine at the time of content composition and are expressing their concerns about taking the vaccine or are directly related to the discussion of side effects experienced by users who have received at least one dose of the vaccine. *Topic 5* appeared to be focused on much broader terms, information (i.e., *news, source, question*) as well as a direct mention of concerns about vaccination. The topic also mentioned *autism,* most likely in reference to the antivaccine movement's fixation on the false narrative that vaccines cause autism. The findings in *Topic 5* are particularly interesting due to the direct detection of discussions related to vac-cine misinformation and autism. For example, a positive-polarity comment sarcastically stated "*Vaccines cause autism, huh? Well, I've got two more days till I get completely upgraded (I get the second Moderna dose on Tuesday)*". While others believed the misinformation such as the content referenced in this negative-polarity post "*MRNA is not safe and should be put into your body unless you are 10000% what it is. MRNA can do/mean literally anything from protecting you from COVID to sterilization all the way into making you autistic (not like the anti-vaxxers thinking the influenza vaccine is the cause for autism-like MRNA can literally be edited to give you mental problems)*". Unfortunately, this post received 28 upvotes, indicating community agreement. These topics generally focused on questions about the vaccine, side effects, medication, experiences with the vaccines, and intervals of time.

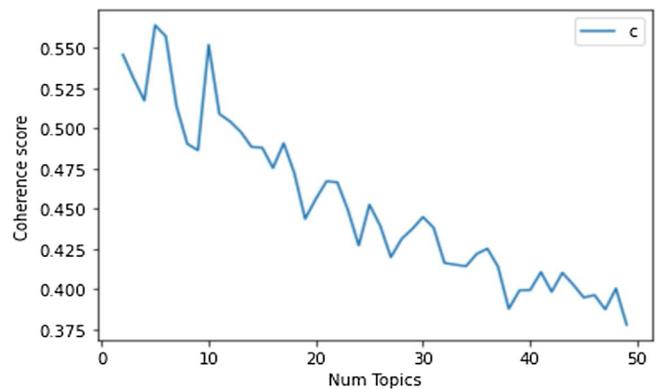

**Fig. 3.** Coherence score vs the number of topics. The y-axis represents coherence and the x-axis represents the number of topics. The coherence score calculation determined 5 Topics were the optimal number for our LDA model.

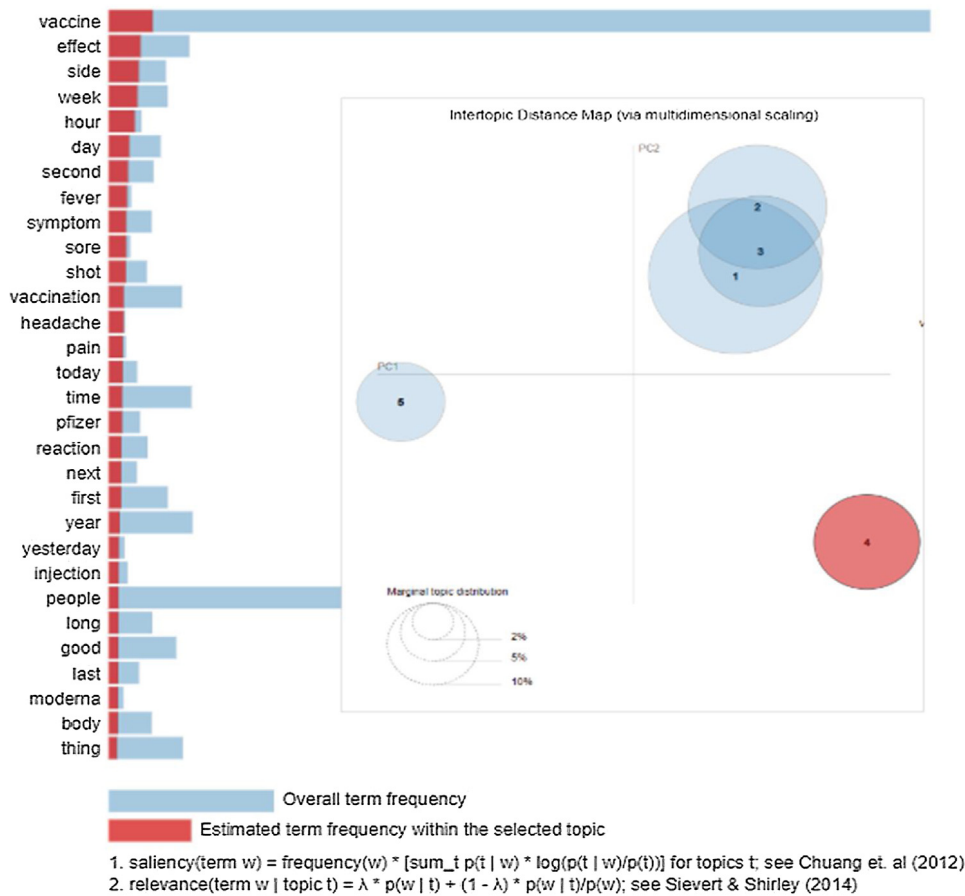

**Fig. 4.** Example of "Topic 4". The blue rectangles are representative of overall term frequency and the red rectangles represent frequency within Topic 1. See github.com/Cheltone/NLP_Reddit for an interactive display of LDA topics. Spheres represent relative topic distribution.

### Monthly analysis

The sentiment analysis results for individual months agreed with the combined analysis and reported that the majority of posts were positive. December reported 57.63% positive, 26.21% negative, and 16.16% neutral. January reported 59.49% positive, 25.52% negatives, and 14.99% neutral. February reported 57.93% positive, 28.10% negative, and 13.97% neutral. March reported 57.13% positive, 25.97% negative, and 16.9% neutral. April reported 54.98% positive, 28.96% negative, and 16.06% neutral. Lastly, May reported the least positive sentiment and highest negative sentiment at 53.05% positive, 30.57% negative, and 16.38% neutral (see GitHub repository).

Overall, the LDA topic modeling results were similar to the complete data set. However, in this case, latent topic quantities were much smaller due to the smaller corpus with each month (less than or equal to three latent topics). The content of these individual months was very similar to the overall combined data set except December. Latent topics in December contain keywords associated with vaccine trials, group, efficacy, as well as potential side effects. January and February both display latent topics related to vaccine dosage, the number of doses, immunity, and side effects. March, April, and May appear to be more closely related. Though these months include topics similarly detected to December through February, these months reference latent topics more directly related to hesitancy (i.e., concern, risk), as well as death. Interestingly, some discussion of T cells was detected in April and May (see Table 1).

### Sentiment topic modeling

The sentiment topic modeling results were significantly more convoluted than the combined and monthly topic model. The models in these three polarities all contain some themes in common related to discussions and questions about the vaccination process, side-effects, concerns, time, and immunity. The negatively classified post topics contained additional keywords such as *government, state, science, employee, risks*, and *several expletives*. Posts classi-fied as neutral displayed topics that referenced *physicians, Pfizer, research, video, link, issue,* and *stories.* Lastly, topics related to the positive posts included terms such as *Moderna, flood, safe, woman, pregnant, family,* and *response.* The positive posts also contained keywords related to death as well as expletives (see Table 2). Table 2 demonstrates LDA topic modeling of the complete dataset from Dec 1, 2020 to May 15, 2021 based on polarity. Please see https://github.com/Cheltone/NLP_Reddit for an interactive topic model.

**Table 1**
Monthly latent topics. It lists two rows of topics and 10 words from each of the combined dataset ranging from Dec 1, 2020 to May 15, 2021, and monthly topics. see https://github.com/Cheltone/NLP_Reddit for an interactive topic model.

|  | Combined data set (December 1, 2020–December 31, 2020) |
|---|---|
| Topic number | Latent topics |
| 1 | vaccine, people, effect, time, many, thing, year, death, month, good |
| 2 | vaccine, effect, side, week, hour, day, second, fever, symptom, sore |
| 3 | vaccine, people, dose, mask, thing, group, datum, year, immunity, efficacy |
| 4 | vaccine, virus, people, immune, system, year, antibody, vaccination, immunity, body |
| 5 | vaccine, question, contact, concern, action, people, source, news, moderator, answer |
| December 2020 |  |
| 1 | vaccine, virus, immune, system, question, cell, protein, infection, symptom, body |
| 2 | vaccine, dose, trial, group, first, efficacy, datum, case, day, participant |
| 3 | vaccine, people, year, thing, effect, time, virus, long, side, good |
| January 2021 |  |
| 1 | vaccine, dose, effect, people, side, day, second, week, first, shot |
| 2 | vaccine, people, virus, year, time, good, immunity, immune, risk, case |
| February 2021 |  |
| 1 | vaccine, dose, second, effect, day, side, week, people, first, hour |
| 2 | vaccine, people, virus, immune, vaccination, immunity, time, antibody, cell, mask |
| March 2021 |  |
| 1 | vaccine, people, virus, mask, year, thing, immunity, good, time, immune |
| 2 | vaccine, vaccination, dose, death, question, effect, week, day, people, concern |
| April 2021 |  |
| 1 | vaccine, people, mask, thing, year, effect, time, vaccination, virus, death |
| 2 | vaccine, people, virus, vaccination, immune, t, immunity, death, effect, time |
| May 2021 |  |
| 1 | vaccine, people, effect, side, time, second, shot, week, death, day |
| 2 | vaccine, people, mask, virus, vaccination, immunity, risk, year, thing, t |

**Table 2**
Latent topics by polarity.

|  | Negative |
|---|---|
| 1 | vaccine, business, people, fucking, fuck, government, mask, health, treatment, free, |
| 2 | vaccine, second, day, side, effect, symptom, dose, shot, week, hour, |
| 3 | vaccine, vaccination, shot, today, itâ€, month, response, day, state, time |
| 4 | vaccine, part, concern, contact, question, appointment, action, resource, employee, helpful, |
| 5 | vaccine, immune, system, body, cold, efficacy, different, variant, cell, term, |
| 6 | vaccine, long, virus, effect, term, people, immune, infection, risk, science, |
| 7 | people, vaccine, vaccination, virus, thing, t, mask, immunity, stupid, sick |
|  | Neutral |
| 1 | vaccine, people, test, antibody, other, part, mask, case, re, body |
| 2 | vaccine, virus, nerve, thing, today, week, physician, doctor, couple, different |
| 3 | vaccine, vaccination, shot, today, itâ€, month, response, day, state, time |
| 4 | people, vaccine, reaction, shot, reason, fever, today, pfizer, vaccination, itâ€ |
| 5 | vaccine, t, work, life, sore, time, different, story, situation, period |
| 6 | vaccine, effect, side, people, second, pfizer, shot, study, year, tomorrow |
| 7 | vaccine, immunity, rate, efficacy, herd, virus, immune, video, issue, link |
| 8 | vaccine, immune, day, cell, system, people, thing, research, site, month |
|  | Positive |
| 1 | vaccine, effect, side, dose, second, day, hour, reaction, death, week |
| 2 | people, vaccine, year, good, thing, time, article, mask, shit, population |
| 3 | immune, virus, system, cell, antibody, body, vaccine, protein, response, immunity |
| 4 | vaccine, doctor, time, trial, right, country, good, link, year, first |
| 5 | vaccine, good, needle, doctor, effective, thing, little, moderna, flood, t |
| 6 | vaccine, people, many, immunity, year, safe, virus, shot, death, effect |
| 7 | vaccine, people, long, time, term, year, virus, thing, many, effect |
| 8 | vaccine, test, risk, trial, woman, infection, pregnant, study, family, people |

## Discussion

### Interpretation

Although the results displayed in this document suggest that public sentiment in Reddit communities is overall positive regarding discussions about the Covid-19 vaccine or experiences with taking the vaccine, keywords and topics were detected that indicate some hesitancy amongst these users. Our results report a higher positive polarity in general, but they do not suggest that the sentiment of these community members has changed significantly during the time interval in focus. This occurrence could be due to the potential bias in these communities and/or related to strict Reddit community guidelines that result in the removal of certain posts, creating either an evidence-based or nonevidence-based echo chamber. It is conceivable that bias could be lessened by amalgamating comments from a right-leaning, left-leaning, and neutral news organization from multiple social media platforms simultaneously [15]. Moreover, it is possible that the sentiment analysis reflected the nature of interaction between users rather than actual feelings about vaccination. Qualitative analysis revealed the detection of some comments that expressed a negative sentiment of the vaccine but were given a positive polarity due to certain aspects in the text. For example, the comment, "*Looking forward to being treated like the plague for refusing the – gene-therapy – vaccine. As a proud introvert, I can't wait for people to avoid me!*", received a polarity score of 1. Nonetheless, these results shed light on user activity within these subreddits and suggest that most active community members participate mainly through the upvote/downvote feature. This behavior is demonstrated by the large discrepancy in authors (∼9000) compared to comment upvotes (612,217), not to mention the other 4.9 million community members who mainly consume the content without interacting.

Topic modeling quality is often challenging to evaluate because using coherency and perplexity are based on purely numerical relationships in word occurrences. At times, an optimal coherence value may result in topics that are not qualitatively coherent [29]. Due to this fact, it is fundamentally necessary to inspect returned topics as well as data content. In our study, qualitative analysis more or less agreed with coherence values. LDA results presented in these models appeared to keep a common theme over time when considering the month-to-month analysis. Moreover, slight changes in portions of topics are still observable that reflect an evolution in discussion from the early vaccine rollout to vaccines being commonly available. Significantly, one constant topic that was detected throughout each month, regardless of polarity is side effects. This finding was expected considering many recently vaccinated people discuss and compare side effects on social media as well as in person. Due to the severity of some documented side effects and their wide media coverage, it's highly conceivable that side-effects are a major contributor to hesitancy. It is also mentionable that the majority of conspiracy theories were not detectable by the LDA models, indicating a minimal occurrence. Besides the mention of *autism* most manually read conspiracy theories were sarcastic (e.g., My nanobots must not be working cause my 5G sucks).

*Limitations*

Our study has some limitations. Additional challenges occur when conducting sentiment analysis in social media text due to long-standing problems with detecting sarcasm, often leading to false positives or false negatives. At least one false positive was detected in a comment thread. One user posted sarcastically, ¨*They never gave me a bloody sticker!*¨ Though a human can see the sarcastic intent of such a post, TextBlob rated this post as negative and highly subjective. Moreover, TextBlob usually exhibits modest returns inaccuracy (50–70%), and there may be room for improvement.

Reddit is superior to other social media platforms in several ways in user numbers and data quality. Though many outstanding studies have been conducted using Twitter data, it is estimated that approximately 50% of Twitter accounts could be BOTS [30]. A recent study by Memon and Carly (2020) reported that up to 14% COVID-19 related posts on Twitter were composed by BOTS [31]. Though some BOTS exist, the operational design of Reddit community interaction does not lend itself to typical BOT behavior. Nonetheless, the site still is not perfect. Only broad data are available regarding the Reddit userbase. While some demographic, financial, gender and geographic data have been gathered [32], geotagged posts are not a regular occurrence on most Reddit posts. Moreover, high-resolution demographic data are not available or recorded. This lack of geocoded data makes comparison with specific regional/city-wide polling or surveys impossible unless the subreddit is explicitly based on a geographical community or dedicated to a specific demographic. That being, Reddit data are typically not ideal for studies of a highly specific geographic area or demographic studies.

*Social media and digital health technologies*

Alongside the numerous public health preventive measures (i.e., social distancing, shelter-in-place, stay-at-home orders, lockdowns, quarantine, etc.) implemented to control the spread of the virus, there is general scientific consensus that the COVID-19 vaccine is protective against the SARS-COV-2. However, the spread of misinformation, disinformation, and fake news plays a significant role in vaccine hesitancy, low vaccination rates, disease outbreaks as well as morbidities and untimely deaths from vaccine-preventable illnesses. Accordingly, the leverage of textual data obtained from social media platforms could facilitate rapid *and* inexpensive public sentiment analysis thereby enabling the implementation of appropriate messaging, digital interventions, and policies. Digital health technologies and Artificial Intelligence [35] are novel, ideal, and effective tools that could facilitate the delivery of accurate, timely, and targeted health information to the general public. For instance, this intervention could be implemented as automated personalized messages and education delivered to individuals based on the content and sentiments from their social media posts. High-impact personalized educational interventions providing clear, unambiguous recommendations/policies/messages on vaccine safety, efficacy, availability, accessibility, affordability, and acceptability, etc. could be impactful. Pivoting online forum discussions on vaccines to accurate and evidence-based information would conceivably facilitate Precision Health Promotion [36] and increased health literacy to promote vaccine confidence.

**Conclusion**

Analysis of a Covid-19 vaccine-related content from 13 subreddits suggests that the sentiments expressed in these social media communities are overall more positive than negative but have not meaningfully changed since December 2020. Nonetheless, keywords indicating vaccine hesitancy were detected throughout the LDA topic modeling. Though this study offers some insight into the public mind, additional work research is still needed to fully understand how to reach populations who feel negative towards the Covid-19 vaccine, and to combat misinformation. The results we present here are the first of an ongoing study to explore vaccine-related content on social media with a focus on identifying and combating misinformation. Future work will investigate these phenomena further by employing dynamic topic modeling with deep learning, semantic networks [33], and other machine/deep learning techniques to develop an optimal system to identify misinformation and intervene [34] within social media. Future work will also involve annotating approximately 20% of our data set with the intent to incorporate supervised machine learning and deep learning techniques for future analyses. Topic modeling could be used to analyze a wider variety of these data sources and could contribute to an even more realistic representation of population sentiment.